\documentstyle[amssymb,secnumtab]{fbssuppl}

\input epsf

\title{Perturbative QCD Applied to Baryons}
\author{Carl E. Carlson\thanks{{\it e-mail address:} 
carlson@physics.wm.edu}}
\institute{Nuclear and Particle Theory Group,
     Department of Physics,\\
     College of William and Mary,
     Williamsburg, VA 23187-8795, USA}

\begin{document}

\maketitle
\begin{abstract}

We review standard applications of perturbative QCD to baryon production,
and argue by examining data that it is generally relevant at
high but experimentally feasible momentum transfers.  Then we consider
some new initiatives, particularly meson photoproduction off baryons and
the seeming quagmire of $\Delta(1232)$ electroproduction.

\end{abstract}

\vglue -10 cm
\noindent {\small \bf WM-98-116}
\vglue 10.6 cm


\section{Introduction}


This talk is a special one at a workshop dedicated to nonperturbative
methods in baryon physics.  It discusses the other side of
things, namely perturbative QCD (pQCD) applied to baryons, with particular
emphasis on applications to exclusive and semi-exclusive reactions.

We will start out in the next section discussing what I will call ``standard
old stuff,''  reviewing methods of calculation and scaling and normalization
predictions that are well known to many, and seeing in what kinematic regime
pQCD seems to work and how well it works there.  I might say now that I am
an optimist, thinking that pQCD results can be valid when momentum transfers
are only a few GeV.  The ``standard old stuff'' will come in three headings,
namely the scaling behavior expected for amplitudes at high momentum
transfer, with comparison to data, the polarization behavior expected for
amplitudes at high momentum transfer, with comparison to data, and some
review of results that have been gotten in the few cases where normalized
calculations are possible.

To balance the old,  section~\ref{newstuff} will present a selection of new
initiatives using pQCD, focusing on semi-exclusive reactions and
connections between low and high momentum transfer behavior of
$\Delta(1232)$ electroproduction.




\section{Standard Old Stuff}


\subsection{Scaling---expectations and data}

Perturbative QCD for exclusive reactions~\cite{bl80} begins by drawing all
the relevant lowest order Feynman diagrams.  There can be many for a given
process and calculating all of them can be time consuming.  However, the
scaling behavior is generally the same for all the diagrams, and can be
ferreted out relatively easily.  The general categories of processes are
form factors at high momentum transfer,  or quasi-elastic reactions at
high $s$ at fixed large $\theta_{CM}$.  An example of the latter,
specifically for
$\gamma p \rightarrow \pi^+ n$, is given in the Figure below.  The
momentum transfer dependence comes from the internal propagators---a
$1/Q^2$ for each gluon propagator (where $Q$ is some momentum scale) and
a $1/Q$ quark propagators---and a factor $Q$ for each quark
line~\cite{cg84,pire}.

\begin{figure} [h]        \label{piphoto}
\vglue -2.5cm
\hskip 2cm \epsfysize 4cm  \epsfbox{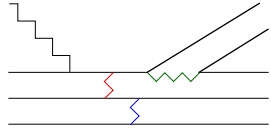} 

\vglue -2mm
\caption{One lowest order diagram for $\gamma p \rightarrow \pi^+ n$.}
\end{figure}

The amplitude represented by this diagram has four quark lines and three
each of internal quark and gluon propagators.  Hence
\begin{equation}
{\cal M} \propto Q^4 Q^{-3} (Q^2)^{-3} = Q^{-5} \propto s^{-5/2},
\end{equation}

\noindent and the differential cross section is 
\begin{equation}
{d \sigma \over dt} = {1\over 16 \pi s^2} |{\cal M}|^2 \propto s^7.
\end{equation}

Does it work? Here is a plot of $s^7 d\sigma / dt$ vs. $s$ for
$\theta_{CM}=90^\circ$,

\begin{figure} [h]   \label{jimmypi}
\centerline{ \epsfysize 4.7cm \epsfbox{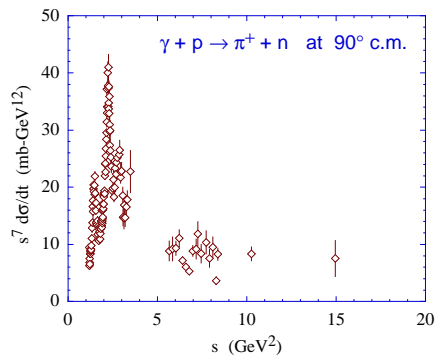} }

\vglue -2mm
\caption{Scaled cross section for $\gamma p \rightarrow \pi^+ n$.}

\end{figure}

\noindent
The bumps at low $s$ are resonance excitations, and the pQCD
expectation appears to succeed just above resonance region.

Form factors for electron elastic or quasi-elastic scattering from a
hadron with $N$ constituents generally go like,
\begin{equation}  \label{falloff}
F(Q^2) \propto 1/(Q^2)^{N-1}.
\end{equation}

\noindent For baryon elastic or transition form factors this means 
$F \propto 1/Q^4$.  (At least the leading form factor falls like this:  there
may be form factors that are zero to leading order, which then fall faster.)

Paul Stoler~\cite{stoler} has produced the following plots:  

\begin{figure} [h]
\centerline{\epsfysize 3 cm \epsfbox{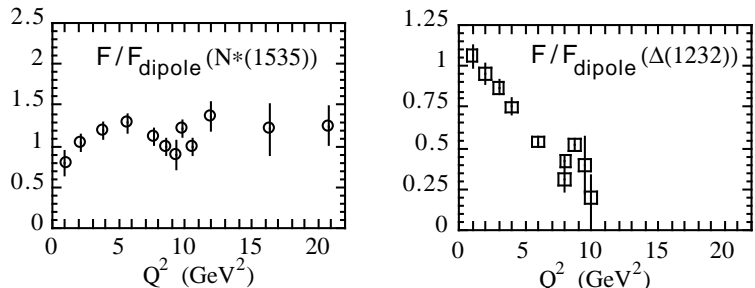}}
\label{stoler}

\vglue 3.5mm
\caption{Form factors for two transition form factors, divided by 
$F_{\rm dipole}$}
\end{figure}

\vglue -3mm
\noindent For reasons of space, we have shown only the nucleon
to $N(1535)$ and to $\Delta(1232)$ transition form factors. The dipole form
is $(1+Q^2/0.71 {\rm GeV}^2)^{-2}$, so a flat curve is what pQCD predicts.
There are also plots for the elastic case and the $N(1688)$ region, which
look rather like the $N(1535)$.  Hence the pQCD results are
successful, except for the $\Delta(1232)$.  

The $\Delta(1232)$ falls faster than the others.  There is a
reason within the pQCD framework for this and a discussion will come in
section~\ref{ddr}.  Also, there has been a suggestion
that the $N(1535)$ is a $\Lambda K$ bound state.  This makes the
minimum Fock component a 5 constituent state, with a faster form factor
falloff according to Eqn~(\ref{falloff}). This is
not supported by the data.
        
          
\subsection{Polarization---expectations and data}      \label{helicity}


The scaling rules tell us the leading scaling behavior, assuming nothing
else suppresses the amplitude farther.  In particular, there can be
farther suppression if the helicity conservation rules are violated. 
The basic rule is that, neglecting quark mass and binding, the quark
helicity is conserved in interactions with either gluons or
photons.  If all interactions are at close range, the orbital
angular momentum of the quarks can be neglected, and then the
helicity of the hadrons overall must be conserved.  Each unit violation
of the helicity conservation rule costs a factor of $O(m/Q)$ where $m$ is
some mass scale and $Q$ is some momentum transfer scale~\cite{cg84,pire}.

The nucleon electromagnetic form factors give a simple example.  Thinking
in the Breit frame, a transverse photon with helicity $+1$ hitting a
nucleon with helicity $+1/2$ gives a final state nucleon also of helicity
$+1/2$.  Hadron helicity is conserved;  The previous rules apply.  The
result in terms of $G_M$ comes from  
\begin{equation}
G_+ = {1\over 2m_N} \langle R, \lambda^\prime = {1\over 2} |
   \epsilon_\mu^{(+)} \cdot j^\mu(0) | N, \lambda = {1\over 2} \rangle
   =  {Q\over m_N \sqrt{2}}  G_M \propto {1\over Q^3}
\end{equation}

\noindent and so one gets $G_M \propto 1/Q^4$, which is well known to
be true.  However, bringing in a longitudinal photon leads to a final
helicity of $-1/2$, and so the amplitude should be suppressed by  a power
of $Q$, and
\begin{equation}
G_0 = {1\over 2m_N} \langle R, \lambda^\prime = {1\over 2} |
   \epsilon_\mu^{(0)} \cdot j^\mu(0) | N, \lambda = {1\over 2} \rangle
   =    G_E \propto {1\over Q^4}   .
\end{equation}

Thus for the Pauli
form factor $F_2$ (using $\tau \equiv Q^2/4m_N^2$),
\begin{equation}
F_2 = {G_M + G_E \over 1 + \tau} \propto {1 \over Q^6}.
\end{equation}

\noindent Comparing to $F_1$ in the figure 
($F_1 = G_E + \tau G_M / (1 + \tau) \propto 1/Q^4$), one
sees that this prediction from hadron helicity conservation proves to be
true in nature~\cite{bosted}.

\begin{figure} [h]
\centerline{ \epsfysize 4.5cm \epsfbox{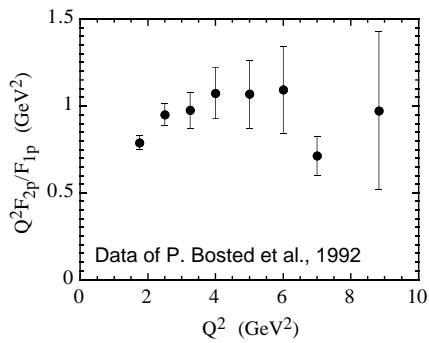} }
\label{bosted92}
\caption{Checking the $F_2$ scaling behavior vs. data.}
\end{figure}


\subsection{Normalized calculations}         \label{ddr}


When normalized calculations can be done, they become the heart of the
perturbative predictions for exclusive reactions.  For example, for some
typical form factor the whole high momentum transfer calculation is
\begin{equation}
F(Q^2)=  \int [dx] [dy] \phi(x,Q^2) T(x,y,Q^2) \phi(y,Q^2)    
\end{equation}

\noindent Here $\phi(x)$ is the distribution amplitude for the final
baryon, simply related to its wave function, and describes finding three
quarks with substantially parallel momenta, with a tolerance related to
the scale $Q	$, and with momentum fractions $x_i$; $\phi(y)$ is the same
for the initial state.   The distribution amplitudes are only weakly
dependent on $Q$.  The main, power law, $Q$ dependence comes from the
amplitude $T$, which describes one quark absorbing a large momentum
transfer 
$Q$ and sharing it with the other quarks so they are all parallel moving
in the final state.  It is calculated in perturbation theory.  

The wave function
or distribution amplitudes cannot be calculated in perturbation
theory.  One gets them using QCD
sum rules to get moments of wave functions, which become constraints on
model wave functions, and model wave functions have been offered by, for
the nucleon, CZ and COZ	(Chernyak, Oglublin, Zhitnitsky) and	KS
(King-Sachrajda) and  GS			 (Gari-Stefanis).

These all lead to good results for proton $G_M$ (of course),
\begin{equation} 
Q^3 G_+(p \rightarrow p) \approx 0.75 {\rm \ GeV}^3  ,
\end{equation}

\noindent with 

\begin{equation}  \label{asymp} 
Q^3 G_+(N \rightarrow \Delta) \approx 0.08 {\rm\ GeV}^{3}
\end{equation}		 

\noindent and

\begin{equation} 
Q^3 G_+(p \rightarrow N^*(1535)) \approx 0.46 {\rm\ GeV}^{3}  .
\end{equation}		 

\noindent For definiteness, these use KS for the nucleon and CP
(Carlson-Poor) for the $\Delta$ and $S_{11}$ (with
apologies to FOZZ (Farrar, Oglublin, Zhang, Zhitnitsky) and BP
(Bonekamp-Pfeil))~\cite{delta}.

The asymptotic $\Delta$ transition amplitude is small.  Hence what we see
in the data shown earlier is still the subleading part of the transition.
A deep reason not known. Still, we can claim that the DDR (Disappearing
Delta Resonance) is understood within pQCD.

A quick summary of this quick review is that pQCD has a
decent record in explaining data at high
but feasible momentum transfers, for single baryons


\section{New Initiatives}   \label{newstuff}



\subsection{Semi-exclusive reactions}

  
A semi-exclusive reaction is one where one or a few, but not
all, of the hadrons in a final are observed.  We will focus on pion
photoproduction~\cite{many,acw12},
$\gamma p \rightarrow \pi X$.

We will also suppose that the transverse
momentum of the pion is high, and that the recoil mass $m_X$ is
high.  These provisos ensure that perturbation theory can
be used in the calculations.

We hope to learn or supplement what we know about:	

$\bullet$ the polarized and unpolarized gluon distributions of
the target,

$\bullet$ the quark distributions for high x,  and

$\bullet$ the pion wave function at short range.

To proceed, let the transverse momenta be  high  enough (say $k_\perp >
2$ GeV) so that vector meson dominance is a small contributor.  The pion
in 
$\gamma p \rightarrow \pi X$ comes either from a parton emerging in
some direction and fragmenting (so that the pion is part of a jet)
or---at the very highest transverse momenta---directly as part of the
short range process (whence the pion is kinematically isolated).

Where, fragmentation dominates, about 1/3 to 1/2
of rate comes from gluon targets in the proton.  Note the
importance of the high pion transverse momentum, and not just
for allowing perturbative calculations. There has to be a
recoiling particle, hence the process must be higher order. 
Then it is possible for the gluon target process  to be of the same
order of magnitude as a quark target process.

One quantity to consider is 
\begin{equation}
E \equiv A_{LL} \equiv 
   {d\sigma_{R+} - d\sigma_{R-} \over 
      d\sigma_{R+} + d\sigma_{R-}  }  
\end{equation}

\noindent
as a function of $k_\perp$. The $R$ refers to the right handed
polarization of the photon, and the ``$\pm$'' gives the helicity of the
target proton.  The corresponding quantity for the subprocess  
$\gamma g \rightarrow q \bar q$
is $(-)100$\%, so that there is a possibility of great sensitivity
to the gluon polarization.   This is borne out by actual calculations
using a variety of proposed gluon in the proton
distributions~\cite{acw12}.

We will close this section with one more comment.  As lower
energies it is harder to find a fragmentation region between
the direct pion production and VMD regions.  Help may be
available in fishing out gluon target events by looking two jets
or two hadrons $180^\circ$ apart in azimuth angle. Think of
the two parton level diagrams,

\begin{figure}  [ht]
\centerline{\epsfysize .75 in       \epsfbox{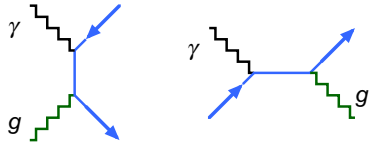}  }

\caption{`Gluon fusion' and `quark Compton' subgraphs for pion
photoproduction.}
\end{figure}

\noindent Fragmenting $q$'s give faster hadrons than fragmenting
glue. Perhaps observing two pions with some cut
like each
$k_\perp$ above  1.5 GeV suffices to ensure that
gluon fusion dominates quark Compton~\cite{twojets} even
at CEBAF with 12 GeV.


\subsection{Approach to pQCD in $\Delta(1232)$ electroproduction}


Electroproduction of the $\Delta(1232)$,
$\gamma^* + N \rightarrow \Delta$,
is a tough place to see pQCD at
work for two reasons.  One is that the low $Q^2$ starting point is so
different from the asymptotic ending point.  In terms of the multipole
amplitudes, the quark model expectation, born out by data, is that
the so-called electromagnetic ratio (EMR) or
$E_{1+}/M_{1+}$ is essentially zero at low $Q^2$, whereas the high $Q^2$
pQCD prediction is that same ratio is unity.  The other is that the
leading term asymptotically is unusually small, as we have already noted
in section~\ref{ddr}.  

Since pQCD seems to work at a few GeV$^2$ in more normal cases,
we~\cite{cm98} thought we should examine how the probably delayed approach
to the pQCD result might go as a function of $Q^2$.  We did so by
choosing simple forms that would give the correct results at low and
high $Q^2$ and that obeyed a few principles.  We worked using the
language of helicity amplitudes, say the $G_+$ and $G_-$ defined in
section~\ref{helicity}.  The principles were basically three:  the
falloffs of $G_+$ and $G_-$ should be $1/Q^3$ and $1/Q^5$
asymptotically;  another is that there should be a kinematic zero in the
amplitude at a (timelike) $Q^2$ where the $\Delta$ does not recoil when
produced off a standing nucleon; and another is the high $Q^2$
normalization (with due regard for the uncertainties of the calculation) 
of
$G_+$ that was quoted in section~\ref{ddr}.

At the photon point, $Q^2=0$, the overall normalization of the two
helicity amplitudes were fixed by comparing to existing data.  The size of
$G_-$, essentially given by the mass parameter governing its falloff in
$Q^2$,  was also determined from unseparated in helicity data on
$\Delta$  electroproduction.  Some tweaking of the $G_+$ mass parameter
was also needed: there was some information about $E_{1+}/M_{1+}$ at 3
GeV$^2$ even before the recent CEBAF data was released.   Results of
our fits are shown in the Figure below.

\begin{figure} [h]
\centerline{\epsfysize 4 cm 
                             \epsfbox{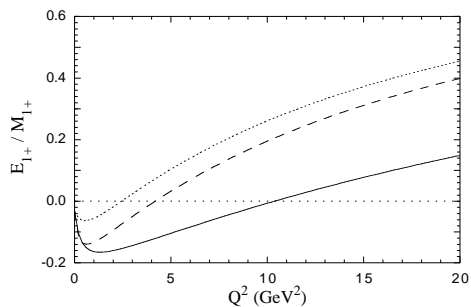}}
\label{emr}
\caption{The electromagnetic ratio for $\Delta$ electroproduction}
\end{figure}

The solid curve is our preferred fit;  the dashed curve is a naive fit
that did not fit the unseparated data well, and the not so different
dotted curve has a asymptotic $G_+$ that was in our opinion too large
even given generous uncertainties in the calculated value.  It appears
that even in this tough situation there will be some push toward the pQCD
result by 10 GeV$^2$ momentum transfer.

\medskip

We have only mentioned two new initiatives because of space and time
limitations.  Others exist, notably~\cite{dvc} the idea of off-forward
parton distributions and applications to deeply virtual Compton
scattering and meson electroproduction and also including new work on
inclusive/exclusive connections.





\begin{acknowledge}

I thank the organizers of this excellent workshop for their hard work,
my collaborators on the projects described in the ``new initiatives
section,'' namely Andrei Afanasev, Nimai Mukhopadhyay, Chris Wahlquist,
and the NSF for support under grant PHY-9600415.

\end{acknowledge}


\makeatletter \if@amssymbols%
\clearpage 
\else\relax\fi\makeatother


\SaveFinalPage
\end{document}